\documentclass[english]{iopart}
\usepackage[T1]{fontenc}
\usepackage[latin1]{inputenc}
\usepackage{graphicx}

\makeatletter

\providecommand{\tabularnewline}{\\}

\usepackage{iopams}
\usepackage{setstack}

\usepackage{setspace}

\usepackage{babel}
\makeatother

\begin{document}
\title[New infrared emission of the NV centre in diamond]{New infrared emission of the NV centre in diamond: Zeeman and uniaxial stress studies}

\author{L J Rogers, S Armstrong, M J Sellars and N B Manson}

\address{Laser Physics Center, RSPhysSE, Australian National University, Canberra,
ACT 0200, Australia}

\ead{lachlan.rogers@anu.edu.au}

\begin{abstract}
An emission band in the infrared is shown to be associated with a
transition within the negative nitrogen-vacancy center in diamond.
The band has a zero-phonon line at 1046 nm, and uniaxial stress and
magnetic field measurements indicate that the emission is associated
with a transition between $^{1}\! E$ and $^{1}\! A_{1}$ singlet
levels. Inter-system crossing to these singlets causes the spin polarisation
that makes the $\mbox{NV}^{-}$ centre attractive for quantum information
processing, and the infrared emission band provides a new avenue for
using the centre in such applications. 
\end{abstract}

\pacs{42.50.Ex, 42.62.Fi, 61.72.jn, 71.70.Ej, 71.70.Fk, 78.30.-j}

\noindent{\it Keywords\/}: {infrared emission, electronic structure, nitrogen-vacancy center}

\maketitle

\section{Introduction}

The electronic structure of the negative nitrogen vacancy ($\mbox{NV}^{-}$)
center in diamond gives it properties that are attractive for aspects
of quantum information applications \cite{Jelezko2004b,Jelezko2004a,Childress2006,Wrachtrup2006,Hanson2006a,Gaebel2006,Santori2006a,Waldermann2007}.
One of the attractive features is the phenomenon of optically induced
spin polarization of the $S=1$ ground state \cite{Loubser1977,Reddy1987,Redman1991}.
It has been proposed that the polarization arises due to inter-system
crossing from the excited triplet state to singlet levels, and decay
back to the ground state with an overall change of spin. This process
was thought to be non-radiative due to the emission intensity dependence
on the level of spin polarisation \cite{Manson2006}, and there was
no direct knowledge of the singlets involved. However, in this work
infrared emission from the $\mbox{NV}^{-}$ centre is reported and
shown to be associated with the singlet levels. Spectral analysis
of this emission has provided information about the polarising decay
path and the electronic structure of the NV centre.

\section{Observations}

\subsection{Emission spectrum }

\begin{figure}
\hspace{71pt}\includegraphics[width=8.6cm]{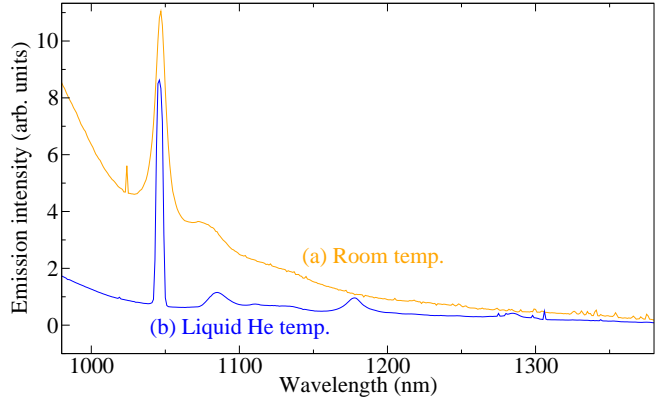}

\caption{Infrared emission band at (a) room and (b) liquid helium temperatures
for the high concentration sample with an $\mbox{NV}^{-}$ center
concentration of $3\times10^{18}\mbox{ cm}^{-3}$. Trace (b) from
the cryogenic measurement has been divided by a factor of 5 on this
graph in order to compensate for the enhanced ZPL intensity. The vibronic
tail of the characteristic $\mbox{NV}^{-}$ visible emission is apparent
under the infrared band. \label{fig:ir-spec-H-hot-cold}}

\end{figure}

\begin{figure}
\hspace{71pt}\includegraphics[width=8.6cm]{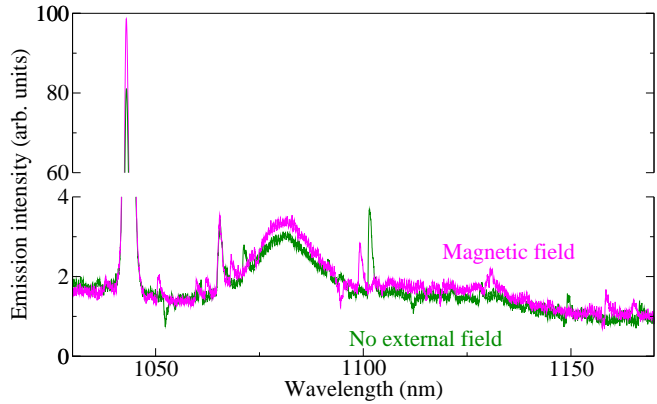}

\caption{Infrared emission band for low concentration ($10^{17}\mbox{ cm}^{-3}$)
sample at liquid helium temperature. The spectrum was measured with
and without an approximately 600 G magnetic field applied, and the
intensity with the field was 122\% of the intensity with no external
field. The sharp feature at 1064 nm common to both traces was due
to some scatter from the 532 nm laser. \label{fig:ir-spec-L-cold}}

\end{figure}

Two synthetic 1b diamond samples of different defect concentrations
were used. Both were 2 mm cubes that had been irradiated and annealed
to produce $\mbox{NV}^{-}$ centers with concentrations of about $3\times10^{18}\mbox{ cm}^{-3}$
({}``high'' concentration) and about $10^{17}\mbox{ cm}^{-3}$ ({}``low''
concentration). Each of the samples had $\langle110\rangle$, $\langle110\rangle$
and $\langle100\rangle$ faces. The samples were excited with a 532
nm laser at 100 mW and the emission at right angles was dispersed
by a monochromator and detected on an ADC model 403L cooled germanium
photodetector. A weak infrared emission band with a zero-phonon line
(ZPL) at 1046 nm was observed, and the spectrum is shown in Figures
\ref{fig:ir-spec-H-hot-cold} and \ref{fig:ir-spec-L-cold}. The characteristic
$\mbox{NV}^{-}$ emission has a higher energy ZPL at 637 nm and vibrational
sidebands that extend beyond 1000 nm, the extreme low energy tail
of which create the intensity baseline in Figure \ref{fig:ir-spec-H-hot-cold}.
Throughout this paper the emission from the 637 nm transition is referred
to as {}``visible'' in order to differentiate it from the infrared
emission band.

At room temperature the infrared ZPL at 1046 nm was clearly discernible
accompanied by a vibrational band, and at low temperature the features
were clearer and dominated by the zero-phonon line. The linewidth
was measured to be 0.3 nm (0.3 meV) in the lower concentrated sample,
but was broadened significantly to 4 nm in the higher concentration
sample. The vibrational sideband had similar integrated area to the
ZPL $(\mathcal{S}\mbox{-coefficient}\approx1)$ and was comprised
of peaks shifted by 42.6 meV (344 $\mbox{cm}^{-1}$), 133 meV (1070
$\mbox{cm}^{-1}$) and 221 meV (1780 $\mbox{cm}^{-1}$) from the ZPL.

\subsection{Magnetic field measurements}

\begin{figure}
\hspace{71pt}\includegraphics[width=8.6cm]{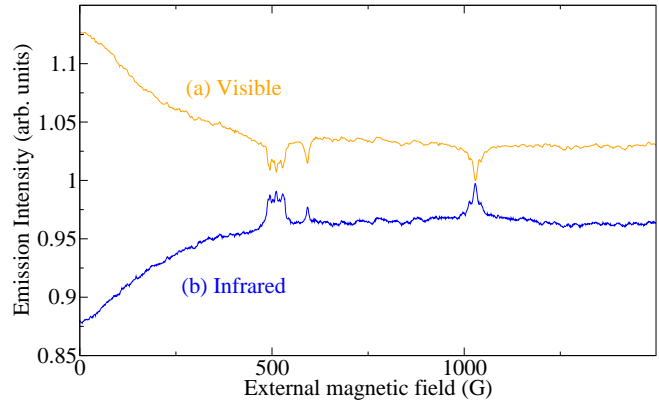}

\caption{Emission intensity for varying external magnetic field. The field
was aligned to a $\langle111\rangle$ crystal axis to within $0.025^{\circ}$.
The magnetic field spectra are shown on the same scale, however the
infrared intensity was many orders of magnitude weaker than that of
the visible. \label{fig:mag_spectra}}

\end{figure}

\begin{figure}
\hspace{71pt}\includegraphics[width=8.6cm]{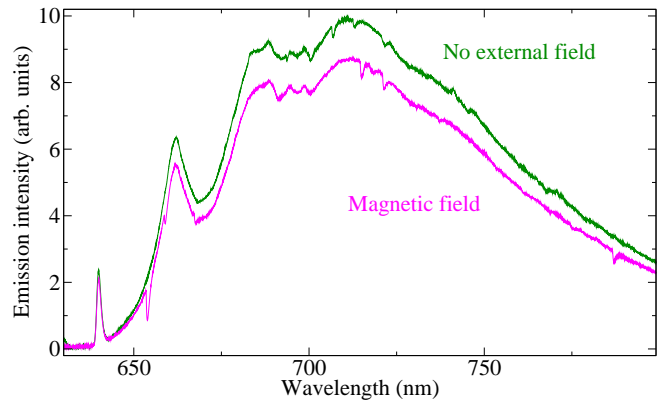}

\caption{Change in the visible emission intensity due to presence of an approximately
600 G external magnetic field. Across the entire band, the intensity
with the field was 87.7\% of the intensity with no magnetic field.
\label{fig:spectrum-vis-500G}}

\end{figure}

The intensity of both infrared and visible emission bands was found
to vary with low magnetic fields, and the variation for the low concentration
sample is shown in Figure \ref{fig:mag_spectra} for fields between
0 and 1500 Gauss. The measurements were made at room temperature,
again involving 100 mW laser excitation at 532 nm. The visible emission
transmitted by a long pass filter at 615 nm was detected by a Si detector,
and the weakness of the infrared emission meant that this signal was
completely dominated by the $^{3}\! A_{2}\leftrightarrow{}^{3}\! E$
transition. The infrared signal was collected through a 1040 nm long
pass filter and detected on an InGaAs detector. In this case the infrared
band dominated, although a small contribution (10 - 15\%) remained
from the vibronic tail of the $^{3}\! A_{2}\leftrightarrow{}^{3}\! E$
transition. Spectrally isolating the emission bands gave a more reliable
measure of the change in emission magnitude with magnetic field, and
this is shown in Figures \ref{fig:ir-spec-L-cold} and \ref{fig:spectrum-vis-500G}.

\begin{figure}
\hspace{71pt}\includegraphics[width=8.6cm]{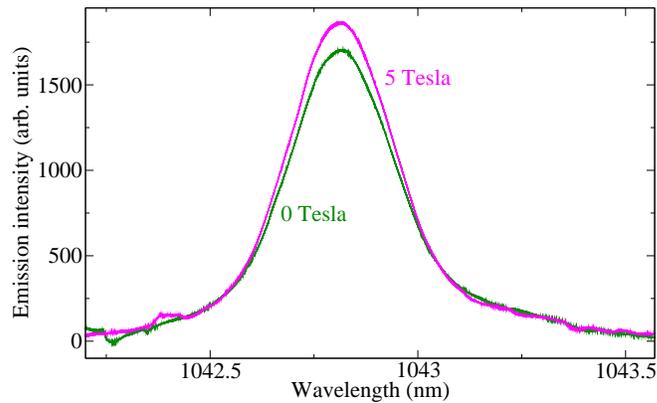}

\caption{Infrared ZPL for 5 T magnetic field and zero field for sample at 4.2
K.\label{fig:ir-zpl-zeeman}}

\end{figure}

High field Zeeman measurements of the 1046 nm line were also undertaken.
These were made for the low concentration sample cooled to 4.2 K where
the linewidth was 0.3 nm. The sample was mounted within the core of
a super-conducting Helmholtz coil, and the magnetic field and 532
nm laser were directed along the $\langle110\rangle$ direction. The
infrared emission was detected at right angles along the $\langle100\rangle$
direction, and the ZPL spectrum for a 5 T field is shown in Figure
\ref{fig:ir-zpl-zeeman}. No Zeeman shift or splitting was observed,
although there was a small change in intensity consistent with the
room temperature measurements.

\subsection{Transient response}

\begin{figure}
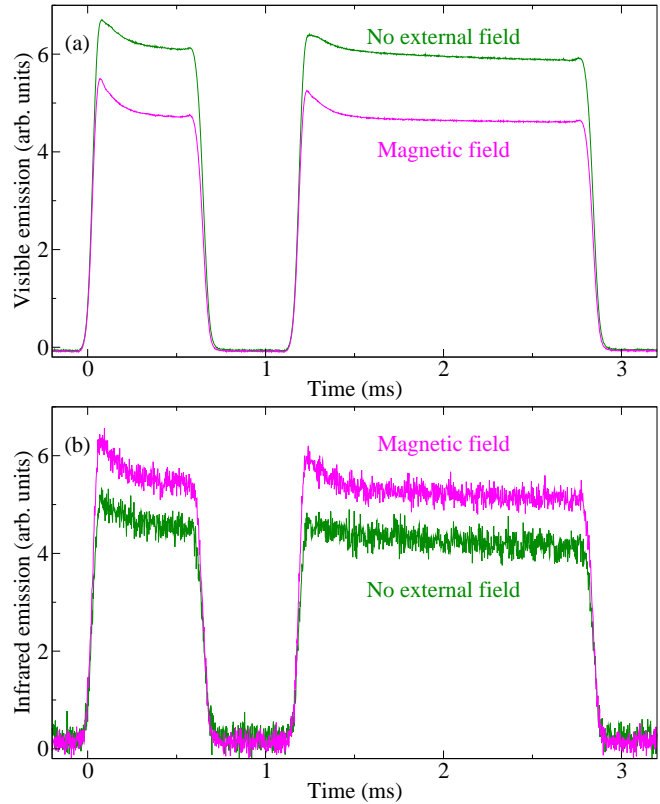

\hspace{71pt}\includegraphics[width=8.6cm]{dynamics_red_slow}

\hspace{71pt}\includegraphics[width=8.6cm]{dynamics_ir_slow}

\caption{Visible (a) and infrared (b) emission for a 700 ns, 1700 ns pulse
pair. Measured with and without a magnetic field of approximately
600 Gauss applied. \label{fig:ir-dynamics-high-intensity}}

\end{figure}

The time dependence of the infrared and visible emission was investigated
for high intensity excitation pulses. Measurements were made with
the sample at room temperature being excited by a focused 532 nm laser
gated by an acousto-optic modulator. The emission was detected using
long pass filters at 615 nm and 1000 nm for the visible and infrared,
respectively, and to ensure consistency between measurements an InGaAs
detector was used in both cases. The pulse sequence consisted of excitation
for 700 ns, followed by a dark delay of 500 ns, and then a slightly
longer second pulse of 1700 ns. The measurements were repeated with
an approximately 600 Gauss field applied to the sample in a random
direction. The responses are shown in Figure \ref{fig:ir-dynamics-high-intensity}.

\subsection{Uniaxial stress}

\begin{figure}
\hspace{71pt}\includegraphics[width=10.5cm]{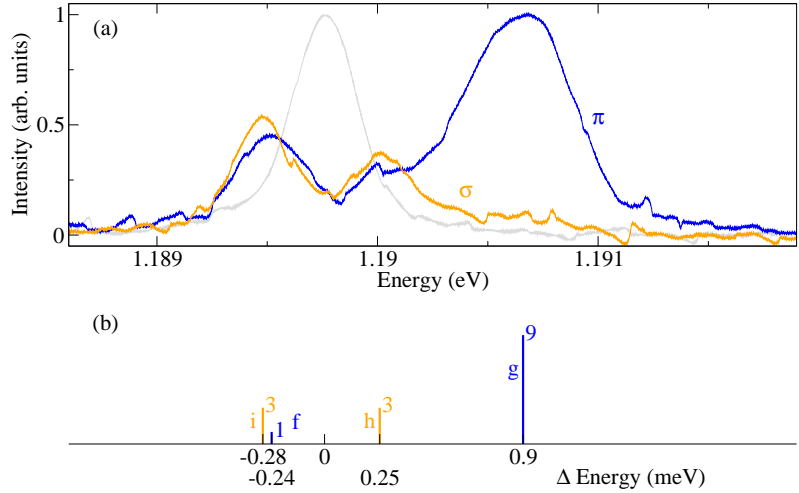}

\caption{(a) Emission spectra for approximately 0.3 GPa stress applied along
the $\langle110\rangle$ direction, measured along the $\langle110\rangle$
direction with polarization parallel $(\pi)$ and perpendicular $(\sigma)$
to stress. The zero stress ZPL is shown in grey for reference. (b)
Theoretical patterns for an $A\leftrightarrow E$ transition at a
site of $C_{3v}$ symmetry, showing the predicted relative intensities
of each line. The splittings are related to the stress coefficients
of \cite{Hughes1967} by $\Delta E_{\mbox{i}}=A_{1}+A_{2}+C-B$; $\Delta E_{\mbox{f}}=A_{1}+A_{2}-C+B$;
$\Delta E_{\mbox{h}}=A_{1}-A_{2}+C+B$; $\Delta E_{\mbox{g}}=A_{1}-A_{2}-C-B$.
Values for these coefficients were obtained from the measured splitting
of the infrared ZPL and are given in Table \ref{tab:Stress-coefficients}.
\label{fig:ir-stress-splitting-spectra-110}}

\end{figure}

\begin{figure}
\hspace{71pt}\includegraphics[width=10.5cm]{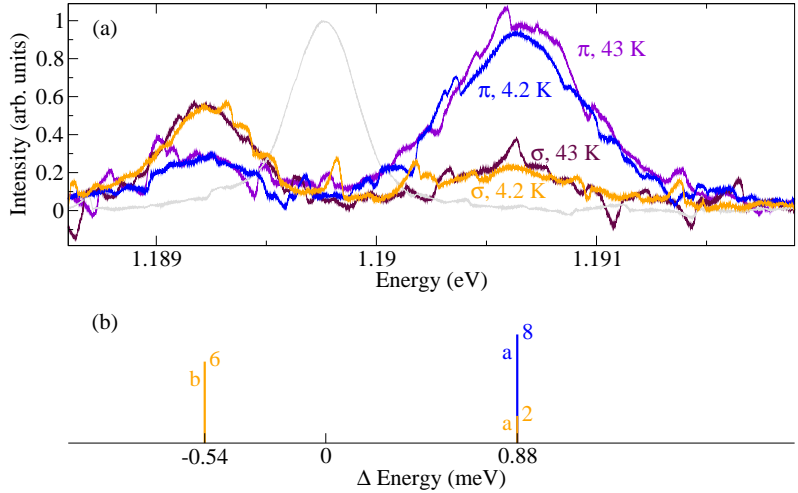}

\caption{(a) Emission spectra for approximately 0.70 GPa stress along $\langle100\rangle$,
measured along the $\langle110\rangle$ direction with polarization
parallel $(\pi)$ and perpendicular $(\sigma)$ to stress. The zero
stress ZPL is shown in grey for reference. (b) Theoretical patterns
predicted for an $A\leftrightarrow E$ transition at a site of $C_{3v}$
symmetry, where the splittings are related to the stress coefficients
of \cite{Hughes1967} by $\Delta E_{\mbox{b}}=A_{1}+2B$ and $\Delta E_{\mbox{a}}=A_{1}-2B$.
\label{fig:ir-stress-splitting-spectra-100}}

\end{figure}

The techniques of uniaxial stress spectroscopy \cite{Kaplyanskii1964,Kaplyanskii1964a,Hughes1967}
were used here to study the zero phonon line at 1046 nm. The low concentration
$2\mbox{ mm}$ cubic sample was held at a temperature of 4.2 K while
being compressed by a piston to pressures up to 0.7 GPa. Excitation
was at 532 nm in the vibronic band of the $^{3}\! A_{2}\leftrightarrow{}^{3}\! E$
transition with polarization parallel to the stress. The emission
at right angles was detected with the polarization in the $\pi$ (parallel
to axis of stress) and $\sigma$ (perpendicular to axis of stress)
directions. Spectra of the 1046 nm zero phonon line were recorded
for stresses along the $\langle110\rangle$ and $\langle100\rangle$
axes, and these are shown in Figures \ref{fig:ir-stress-splitting-spectra-110}
and \ref{fig:ir-stress-splitting-spectra-100}. Three distinct lines
were observed in the spectrum taken for the $\langle110\rangle$ stress
(although the lowest energy line could be two overlapping lines),
and two for the $\langle100\rangle$ stress. The splitting of the
637 nm ZPL was measured, but only used to calibrate the stress and
so is not shown here.

For stress along $\langle100\rangle$, measurements were repeated
at 43 K, and the results are included in Figure \ref{fig:ir-stress-splitting-spectra-100}.
The figure shows that there was no thermal variation in the emission
spectra over this temperature range.

\section{Discussion}

\subsection{Energy scheme for $C_{3v}$}

\begin{figure}
\hspace{71pt}\includegraphics[width=8.6cm]{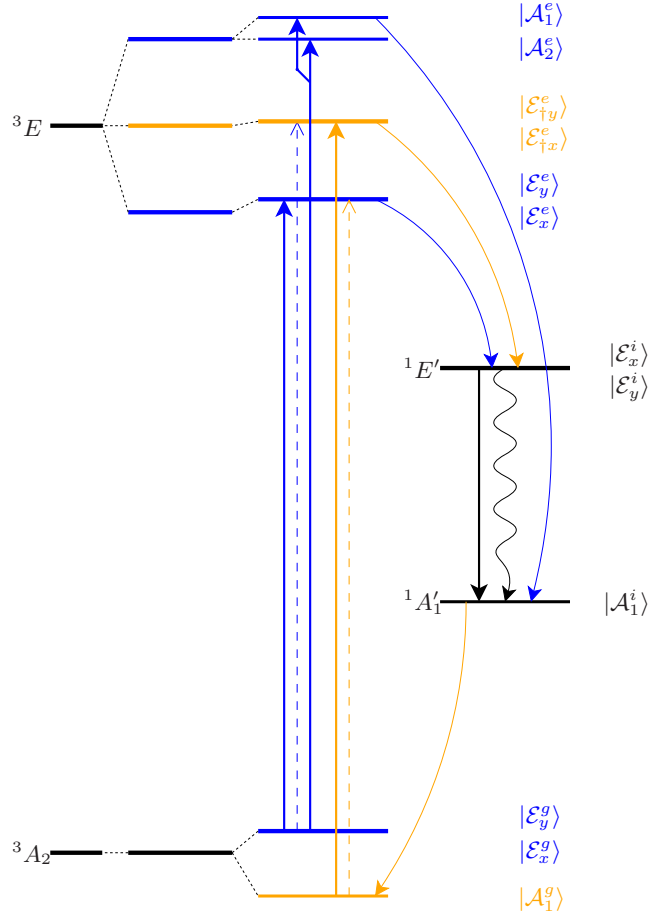}

\caption{The energy levels expected from consideration of the one-electron
states for the six $\mbox{NV}^{-}$ electrons in $C_{3v}$ symmetry
are shown in black (first and fourth columns). Diagonal spin-orbit
terms split the $^{3}\! E$ level as shown in the second column, and
spin-spin interactions give rise to the splittings in the third column.
Straight arrows indicate optical transitions, and the dashed lines
illustrate weakly allowed transitions that prevent perfect spin polarisation.
The curved lines show symmetry-allowed inter-system crossing transitions,
and the wavy line shows suspected vibronic decay between the singlets.
The states are labeled on the left and the symmetry transformation
properties of the spin-orbit wavefunctions are given on the right.
\label{fig:levels-spin-orbit}}

\end{figure}

The low lying electronic states of the NV center have been considered
in previous publications \cite{Manson2006,Manson2007}. The levels
can be obtained by considering the one electron states at the vacancy
site adjacent to the nitrogen. In the notation for $C_{3v}$ point
group symmetry there are two one-electron orbits transforming as an
$A_{1}$ irreducible representation $(a_{1},a_{1}^{\prime})$ and
one transforming as an $E$ irreducible representation $(e)$, and
their energies are considered to be in that order \cite{Lenef1996}.
In the case of the NV$^{-}$ center there are six electrons occupying
these orbits and the lowest energy configuration is $a_{1}^{2}a_{1}^{\prime2}e^{2}$.
This configuration gives rise to $^{3}\! A_{2}$, $^{1}\! A_{1}$,
and $^{1}\! E$ states, whereas there is a higher energy excited triplet
$^{3}\! E$ and singlet $^{1}\! E$ associated with an $a_{1}^{2}a_{1}^{\prime}e^{3}$
configuration. The low lying states are, therefore, as shown in black
in Figure \ref{fig:levels-spin-orbit}. 

The ground state is the $^{3}\! A_{2}$ and the characteristic strong
optical transition at 637 nm is from this ground state to the $^{3}\! E$
excited state. There is fine structure associated with both of the
triplet levels. In the ground state the spin levels are split by spin-spin
interaction into a non-degenerate $M_{s}=0$ level and a degenerate
$M_{s}=\pm1$ level. The excited state is split by diagonal spin-orbit
interaction $\lambda L_{z}S_{z}$ into three equally spaced doubly
degenerate levels and there are small displacements from the non-diagonal
spin-orbit interaction, $\lambda\left(L_{+}S_{-}+L_{-}S_{+}\right)$
\cite{Manson2006}. What is more important for this work is the effect
of spin-orbit interaction between the triplet and singlet levels.
The interaction causes mixing of states transforming as the same irreducible
representation, and states mixed in this way are indicated by curved
arrows in Figure \ref{fig:levels-spin-orbit}. The mixing can enable
radiative or non-radiative transfer between the triplet and singlet
levels, and calculation indicates that in the case of the excited
$^{3}\! E$ state the transfer will be predominantly out of $M_{s}=\pm1$
levels. This is significant as it provides an alternative decay path
to the visible emission. As a consequence the visible emission associated
with $M_{s}=\pm1$ spins is weaker than that for $M_{s}=0$.

\subsection{Magnetic field}

Optical excitation causes preferential population of the $M_{s}=0$
spin projection and the visible emission associated with this spin
is stronger than that for $M_{s}=\pm1$. Thus, as population is transferred
to the $M_{s}=0$ spin state the visible emission increases in intensity
and, conversely, if the spin polarization is reduced the emission
intensity will diminish. A static magnetic field mixes the ground
state spin levels and inhibits population transfer to $M_{s}=0$,
reducing the spin polarisation. Varying the field strength alters
the amount of mixing between levels which changes the intensity of
emission, and this effect is particularly noticeable for an axial
field of 1028 Gauss. At this field value there is a complete mixing
of $M_{s}=-1$ and $M_{s}=0$ states. The population will be equally
distributed between the two spins whereas it will be almost entirely
in $M_{s}=0$ at adjacent magnetic field values. Thus an axially aligned
magnetic field swept through 1028 Gauss causes a marked reduction
in population and noticeable drop in the visible emission intensity
as seen in trace (a) of Figure \ref{fig:mag_spectra}. With the reduction
of spin polarisation there is an increase in the $M_{s}=-1$ population,
which increases the transfer rate to the singlet levels and should
increase the emission intensity from any optical transitions within
the alternative decay path. Exactly such a rise is observed in the
infrared emission at 1028 Gauss. 

All the other intensity variation of the visible emission in Figure
\ref{fig:mag_spectra} can be similarly explained by variation in
the spin polarization of the $\mbox{NV}^{-}$ centre. For example,
at the special cases of axial fields at 514 Gauss and 660 Gauss there
is cross relaxation between the $\mbox{NV}^{-}$ centre aligned with
the field and other spin systems in the crystal (single substitutional
nitrogen defects and non-aligned $\mbox{NV}^{-}$ centres, respectively)
\cite{Holliday1989,Epstein2005}. These other spin systems are not
spin polarised and the cross relaxation will hence reduce the polarisation
of the $\mbox{NV}^{-}$ centre, causing the visible emission to diminish. 

It is immediately apparent from Figure \ref{fig:mag_spectra} that
the infrared emission contains the same features, and two conclusions
can be drawn. Firstly, the complete (anti-) correlation of this intensity
with that of the visible emission (which varies due to $\mbox{NV}^{-}$
spin polarisation) proves the new emission band is associated with
the $\mbox{NV}^{-}$ centre. Secondly, the fact that it is anti-correlated
shows the infrared emission is associated with the population involved
in the inter-system crossing.

\begin{figure}
\hspace{71pt}\includegraphics[width=8.6cm]{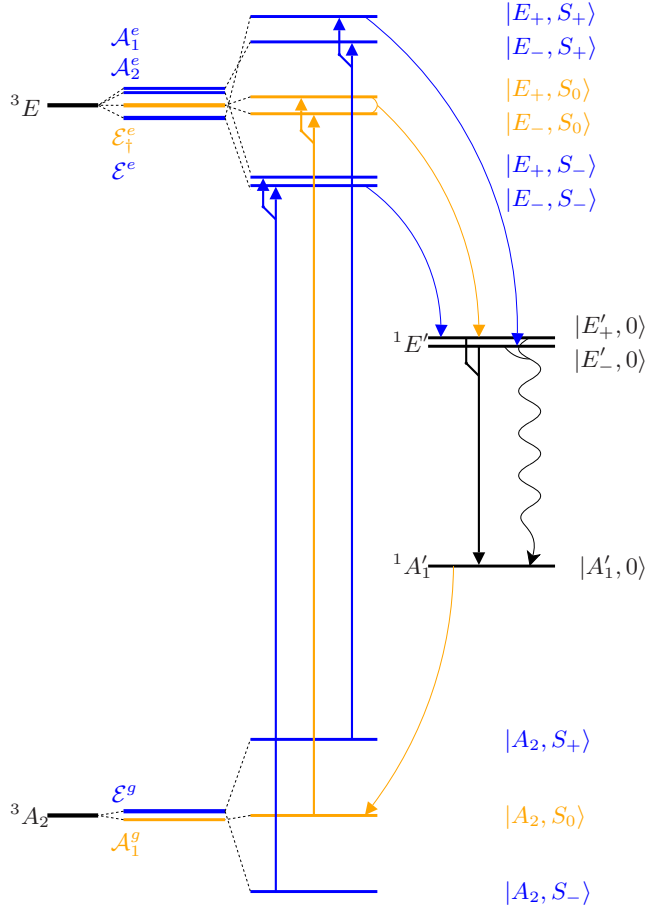}

\caption{Theoretical energy levels for a high (axial) magnetic field situation
are shown in the third column. The first two columns contain the spin-orbit
levels that are shown in detail in Figure \ref{fig:levels-spin-orbit}.
Straight arrows indicate optical transitions, curved lines show symmetry-allowed
inter-system crossing transitions, and the wavy line shows suspected
vibronic decay between the singlets. The wavefunctions in the presence
of a large field are given on the right. \label{fig:Zeeman-splitting.}}

\end{figure}

High magnetic fields were experimentally found not to split the infrared
zero-phonon line. Although this rules out the possibility of the emission
arising from certain transitions, subtleties mean that it does not
conclusively identify the correct transition. A triplet-triplet cannot
immediately be eliminated, as it is already known that the $^{3}\! A_{2}\leftrightarrow{}^{3}\! E$
(637 nm) zero-phonon line is not split by a magnetic field \cite{Reddy1987,Hanzawa1993}.
The individual levels of the triplets are split, but the optical transitions
are between levels of like spin and so they remain degenerate, as
shown in Figure \ref{fig:Zeeman-splitting.}. The same could occur
for the infrared, but this option can be dismissed as there are no
other triplet levels in the $\mbox{NV}^{-}$ system. Unlike triplet
levels, spin-singlets are not split by a magnetic field, and thus
Zeeman splitting of the ZPL would be expected in general for a triplet-singlet
transition. A subtlety here is that no splitting would be observed
if the transition were restricted to a particular spin level of the
triplet state, as is indicated in Figure \ref{fig:Zeeman-splitting.}
for the transition that feeds back to the ground state. However this
diagram is accurate for an axially aligned field only, and the four
different $\mbox{NV}^{-}$ orientations in a bulk diamond sample makes
it impossible to achieve total alignment. Mixing between spin levels
in misaligned centres would lead to observable splitting. 

To first order, no splitting would occur for a singlet-singlet transition.
It is possible for orbitally degenerate states to separate and give
rise to some spectral broadening or splitting, but this can be expected
to be too small to result in a measurable splitting (as it is for
the $^{3}\! A_{2}\leftrightarrow{}^{3}\! E$ transition). The experimental
result in Figure \ref{fig:ir-zpl-zeeman} is most consistent with
the infrared emission arising from the $^{1}\! E\leftrightarrow{}^{1}\! A_{1}$
singlet-singlet transition.

\subsection{Transients\label{sub:Transients}}

The response of the visible emission to intense excitation pulses
has been interpreted previously \cite{Manson2006}. Initially the
$\mbox{NV}^{-}$ centres are evenly distributed between the three
spin projections and they are excited equally. With excitation, the
spin-selective inter-system crossing preferentially populates the
$M_{s}=0$ level and causes spin polarisation as has already been
discussed. This increase in spin polarisation typically increases
the visible emission intensity, however for intense excitation (as
used here) an equilibrium population is built up in a long-lived (300
ns) {}``storage'' state in the singlet system. This decreases the
population that contributes to the visible emission, causing the drop
in visible emission occurring over the first few hundred ns that is
prominent in Figure \ref{fig:ir-dynamics-high-intensity}. At the
start of the second pulse the population is still spin polarised and
so the inter-system crossing is slightly slower. This is observed
as a reduction in the rate that the emission drops to its equilibrium
intensity (ie the storage singlet builds up an equilibrium population).
The peak at the beginning of the second pulse is lower than that of
the first, as some population remains in the storage level after the
500 ns delay and some population is lost through photoionization \cite{Manson2005}. 

The situation is changed by a weak magnetic field, which causes a
mixing of the ground state spin levels and prevents spin polarisation.
As a result, more population takes the alternative decay path through
the singlet levels and a larger equilibrium population is maintained
in the storage state. Thus the visible emission intensity is lower
than it was without the magnetic field. Since the difference between
the first and second pulses was explained by residual spin polarisation,
both pulses should be identical in the case of a magnetic field. In
the experiment, the observed difference that does occur between pulses
is due to imperfect quenching of spin polarisation and also to photoionization
\cite{Manson2005}. 

Comparing the infrared and visible responses to excitation pulses
in Figure \ref{fig:ir-dynamics-high-intensity}, it is immediately
obvious that there is much similarity. The major difference is that
the magnetic field increases the infrared intensity whereas it decreases
the visible, which is consistent with the magnetic field spectra discussed
above. However, the drop in infrared emission intensity within the
first few hundred nanoseconds of each pulse (similar to that obtained
in the visible) indicates that the emitting level is not the {}``storage''
level. This confirms that the infrared band is emitted from the $^{1}\! E\rightarrow{}^{1}\! A_{1}$
singlet-singlet transition, and suggests that the lower singlet state
is the 300 ns storage level. Such a conclusion is plausible, as the
upper singlet could lie close to the excited triplet state to enable
efficient inter-system crossing and the lower singlet could then be
several hundred meV (1000s of $\mbox{cm}^{-1}$) above the ground
state with much slower inter-system crossing. Attempts were made to
re-pump the singlet-singlet transition by exciting in the infrared
to confirm this analysis, but they have not been successful. We were,
therefore, not able to establish where the singlet levels lie in relation
to the triplet levels.

The responses in Figure \ref{fig:ir-dynamics-high-intensity} provide
additional lifetime information. The rates of emission decay from
the initial intensity for each excitation pulse are similar for the
visible and infrared, indicating similar dynamics. The excited state
lifetime for the visible emission is known to be about 12 ns \cite{Collins1983,Batalov2008},
and the signal contrast ratio between spin polarised and unpolarised
(as can be seen in Figures \ref{fig:ir-spec-L-cold} and \ref{fig:spectrum-vis-500G})
indicates that population from the excited triplet crosses to the
singlet system at a similar rate. Thus the population must not spend
extra little time in the upper singlet level before giving the infrared
emission. However, a singlet-singlet transition is unlikely to appreciably
stronger than an allowed triplet-triplet transition and the most likely
explanation for this short lifetime is that there is also competing
non-radiative decay between the singlet levels as indicated by the
wavy line in Figures \ref{fig:levels-spin-orbit} and \ref{fig:Zeeman-splitting.}.
Significant non-radiative decay would account for both the short lifetime
and the weakness of the emission. This may be a general phenomenon
for transitions in the infrared as there are few reports of diamond
emitting at these wavelengths \cite{Zaitsev2001}. Such an increased
contribution of non-radiative decay at low-energy transitions in diamond
can be attributed to the strong electron phonon coupling and high
vibrational frequencies. 

In summary, the following physical description is consistent with
the data. There is an almost 50\% branching of the population from
the $^{3}\! E$ to the singlets, and the upper singlet has a short
lifetime ($<1$ ns) mainly due to non-radiative decay. The lower singlet
level has a longer lifetime ($\approx300$ ns) identified previously
\cite{Manson2006}.

\subsection{Uniaxial stress measurements\label{sub:Uniaxial-stress-measurements}}

The splitting of the zero phonon line with uniaxial stress can be
used to determine the symmetry of the states involved in optical transitions.
A study of this type was undertaken by \cite{Davies1976} and they
showed the 637 nm zero-phonon line to be associated with an $A\leftrightarrow E$
transition at a site of trigonal symmetry. In the present work this
$A\leftrightarrow E$ transition was excited, but energy transferred
within the $\mbox{NV}^{-}$ system gives rise to the infrared transition
which was investigated. 

For stress applied along a $\langle110\rangle$ direction, two pairs
of NV centres have equivalent orientations and both pairs are excited.
In all cases there is a component of strain perpendicular to the $\mbox{NV}^{-}$
axis and for an $A\leftrightarrow E$ transition a maximum of four
lines is therefore predicted \cite{Davies1976,Mohammed1982}. This
is consistent with the experimental observation shown in Figure \ref{fig:ir-stress-splitting-spectra-110},
and the polarization pattern for an $A\leftrightarrow E$ transition
shown below the experimental traces is also in plausible correspondence.
Some deviation of the polarisation pattern is likely to be due to
loss of polarization from scatter from the crystal faces, as they
were not optically polished. 

Parameters for the stress splitting of an $A\leftrightarrow E$ transition
in trigonal symmetry were introduced by \cite{Hughes1967}, and their
values for the present transition were calculated from the energy
splittings of the four lines in Figure \ref{fig:ir-stress-splitting-spectra-110}
and are given in Table \ref{tab:Stress-coefficients}. They are of
the order of a factor 2.5 smaller than those for the 637 nm zero phonon
line determined for by \cite{Davies1976}. The small values for the
strain parameters and the small value for the inhomogeneous linewidth
are consistent with a transition such as $^{1}\! E(a_{1}^{2}a_{1}^{\prime2}e^{2})\leftrightarrow{}^{1}\! A_{1}(a_{1}^{2}a_{1}^{\prime2}e^{2})$,
which involves a spin change but no change of orbit between initial
and final state. 

\begin{table}
\caption{Stress coefficients for the visible and infrared transitions. The
values for the visible are from \cite{Davies1976}, with the signs
of $B$ and $C$ adjusted to reflect the convention in \cite{Mohammed1982}
adopted here. \label{tab:Stress-coefficients}}

\hspace{71pt}\begin{tabular}{|c|c|c|}
\hline 
 & Visible  & Infrared \tabularnewline
 & $(\times10^{-12}\mbox{ eV Pa}^{-1})$ & $(\times10^{-12}\mbox{ eV Pa}^{-1})$\tabularnewline
\hline
\hline 
$A_{1}$ & $1.47$ & $0.53$\tabularnewline
\hline 
$A_{2}$ & $-3.85$ & $-1.44$\tabularnewline
\hline 
$B$ & $-1.04$ & $-0.51$\tabularnewline
\hline 
$C$ & $-1.69$ & $-0.58$\tabularnewline
\hline
\end{tabular}
\end{table}

For a stress along the $\langle100\rangle$ direction all four $\mbox{NV}^{-}$
orientations are at the same angle to the stress and the excitation
polarisation. Each orientation is thus equally excited, and for an
$E\rightarrow A$ transition the component of strain perpendicular
to the $\mbox{NV}^{-}$ axes would lift the degeneracy of the excited
$E$ state. The splitting would be the same for all orientations,
and so produce two lines. The spectra for $\langle100\rangle$ stress
shown in Figure \ref{fig:ir-stress-splitting-spectra-100} is consistent
with this description, and there is plausible correspondence with
the expected polarization pattern shown below the experimental traces.

Since the spin polarisation is to the $M_{s}=0$ level of the ground
state, and this spin level transforms with $A_{1}$ symmetry as indicated
in Figure \ref{fig:levels-spin-orbit}, the $^{1}\! A_{1}$ state
should be the lower of the singlet states \cite{Manson2006}. With
the $^{1}\! E$ as the upper level, the Boltzmann population distribution
of the split components might be expected to change with temperature.
This would cause a change in the relative intensities of the lines
in the spectrum, but no such change was observed between a temperature
of 4.2 and 43 K (Figure \ref{fig:ir-stress-splitting-spectra-100}).
In Section \ref{sub:Transients} it was argued that the lifetime of
this upper singlet is very short, and it is possible that there is
insufficient time to establish a Boltzmann distribution.

\section{Conclusion}

An emission band has been observed in the infrared, with a zero phonon
line at 1046 nm. Measurements have established that this infrared
emission is associated with the $\mbox{NV}^{-}$ defect centre, which
has previously been investigated through its well documented visible
transition at 637nm. From theoretical considerations, magnetic field
and uniaxial stress measurements the infrared emission is attributed
to a $^{1}\! A_{1}\leftrightarrow{}^{1}\! E$ transition where these
singlet levels lie between the ground and excited state triplets.
Although the results presented here are consistent with the $^{1}\! E$
being the higher of the singlets, the order of the levels has not
been conclusively established. Some contention over the order of these
singlet levels already exists due to previous numerical calculations
\cite{Goss1996,Gali2008}.

There are some further puzzling features about the infrared emission
that remain unresolved. The documented energies of the vibrational
sidebands associated with the visible transition (66 and 140 meV \cite{Davies1974})
are not matched by the infrared sideband energies (42.6, 133 and 221
meV). The first of these is very small and the last is large, well
outside the range of single phonons in diamond. The origin of these
frequencies is unclear but the small electron-phonon coupling, as
well as the small stress parameters and inhomogeneous line width,
are consistent with theoretical models of a transition involving only
a spin reorientation. Another strange aspect is the weakness of the
infrared emission. The inter-system crossing branching ratio is as
high as 50\% for the $M_{s}=\pm1$ spin state but yet the infrared
emission is orders of magnitude weaker than that of the visible emission.
The explanation that has been advanced is that there is also very
significant non-radiative decay for the same transition. Related to
this is that with emission arising almost exclusively from $M_{s}=\pm1$
spins then it is surprising that it only drops 15\% with spin polarization.
It suggests that the level of spin polarization attained in our the
samples was very small. 

Despite these issues, the observation of the singlet to singlet transitions
adds significantly to our understanding of the electronic structure
of the NV centre. It provides a new avenue whereby the centre, and
in particular the process of spin polarisation, can be studied and
used for quantum information processing applications.

\ack{}{This work has been supported by Australian Research Council. }

\section*{References}{}

\bibliographystyle{unsrt}
\bibliography{/home/lachlan/phd/reference/resources}

\begin{thebibliography}{10}

\bibitem{Jelezko2004b}
F~Jelezko and J~Wrachtrup.
\newblock Read-out of single spins by optical spectroscopy.
\newblock {\em J. Phys. Condens. Matter}, 16:R1089--R1104, 2004.

\bibitem{Jelezko2004a}
F.~Jelezko, T.~Gaebel, I.~Popa, M.~Domhan, A.~Gruber, and J.~Wrachtrup.
\newblock Observation of coherent oscillation of a single nuclear spin and
  realization of a two-qubit conditional quantum gate.
\newblock {\em Phys. Rev. Lett.}, 93(13):130501, 2004.

\bibitem{Childress2006}
L.~Childress, M.~V.~Gurudev Dutt, J.~M. Taylor, A.~S. Zibrov, F.~Jelezko,
  J.~Wrachtrup, P.~R. Hemmer, , and M.~D. Lukin.
\newblock Coherent dynamics of coupled electron and nuclear spin qubits in
  diamond.
\newblock {\em Science}, 314:281, 2006.

\bibitem{Wrachtrup2006}
J\"org Wrachtrup and Fedor Jelezko.
\newblock Processing quantum information in diamond.
\newblock {\em J. Phys. Condens. Matter}, 18:S807--S824, 2006.

\bibitem{Hanson2006a}
R.~Hanson, O.~Gywat, and D.~D. Awschalom.
\newblock Room-temperature manipulation and decoherence of a single spin in
  diamond.
\newblock {\em Phys. Rev. B}, 74(16):161203, 2006.

\bibitem{Gaebel2006}
Torsten Gaebel, Michael Domhan, Iulian Popa, Christoffer Wittmann, Philipp
  Neumann, Fedor Jelezko, James~R. Rabeau, Nikolas Stavrias, Andrew~D.
  Greentree, Steven Prawer, Jan Meijer, Jason Twamley, Philip~R. Hemmer, and
  Jorg Wrachtrup.
\newblock Room-temperature coherent coupling of single spins in diamond.
\newblock {\em Nature Physics}, 2:408, June 2006.

\bibitem{Santori2006a}
C.~Santori, D.~Fattal, S.~M. Spillane, M.~Fiorentino, R.~G. Beausoleil, A.~D.
  Greentree, P.~Olivero, M.~Draganski, J.~R. Rabeau, P.~Reichart, B.~C. Gibson,
  S.~Rubanov, D.~N. Jamieson, , and S.~Prawer.
\newblock Coherent population trapping in diamond n-v centers at zero magnetic
  field.
\newblock {\em Opt. Express}, 14:7986--7993, 2006.

\bibitem{Waldermann2007}
F.C. Waldermann, P.~Olivero, J.~Nunn, K.~Surmacz, Z.Y. Wang, D.~Jaksch, R.A.
  Taylor, I.A. Walmsley, M.~Draganski, P.~Reichart, A.D. Greentree, D.N.
  Jamieson, and S.~Prawer.
\newblock Creating diamond color centers for quantum optical applications.
\newblock {\em Diamond Relat. Mater.}, 16:1887 -- 1895, November 2007.

\bibitem{Loubser1977}
J.H.N. Loubser and J.A.~Van Wyk.
\newblock Optical spin-polarisation in a triplet state in irrdiated and
  annealed type 1b diamonds.
\newblock {\em Diamond Res.}, 1:11--15, 1977.

\bibitem{Reddy1987}
N.~R.~S. Reddy, N.~B. Manson, and E.~R. Krausz.
\newblock Two-laser spectral hole burning in a colour centre in diamond.
\newblock {\em J. Lumin.}, 38:46, December 1987.

\bibitem{Redman1991}
D.~A. Redman, S.~Brown, R.~H. Sands, and S.~C. Rand.
\newblock Spin dynamics and electronic states of n-v centers in diamond by epr
  and four-wave-mixing spectroscopy.
\newblock {\em Phys. Rev. Lett.}, 67(24):3420--3423, Dec 1991.

\bibitem{Manson2006}
N.~B. Manson, J.~P. Harrison, and M.~J. Sellars.
\newblock Nitrogen-vacancy center in diamond: Model of the electronic structure
  and associated dynamics.
\newblock {\em Phys. Rev. B}, 74(10):104303, 2006.

\bibitem{Kaplyanskii1964}
A~A Kaplyanskii.
\newblock {\em Opt. Spectrosc.}, 16:329--337, 1964.

\bibitem{Kaplyanskii1964a}
A~A Kaplyanskii.
\newblock {\em Opt. Spectrosc.}, 16:557--565, 1964.

\bibitem{Hughes1967}
A~E Hughes and W~A Runciman.
\newblock Uniaxial stress splitting of doubly degenerate states of tetragonal
  and trigonal centres in cubic crystals.
\newblock {\em Proc. Phys. Soc.}, 90:827--838, 1967.

\bibitem{Manson2007}
N~B Manson and R~L McMurtrie.
\newblock Issues concerning the nitrogen-vacancy center in diamond.
\newblock {\em J. Lumin.}, 127:98--103, 2007.

\bibitem{Lenef1996}
A.~Lenef and S.~C. Rand.
\newblock Electronic structure of the n-v center in diamond: Theory.
\newblock {\em Phys. Rev. B}, 53(20):13441--13455, May 1996.

\bibitem{Holliday1989}
K~Holliday, N~B Manson, M~Glasbeek, and E~van Oort.
\newblock Optical hole-bleaching by level anti-crossing and cross relaxation in
  the n-v centre in diamond.
\newblock {\em J. Phys. Condens. Matter}, 1(39):7093--7102, 1989.

\bibitem{Epstein2005}
R.~J. Epstein, F.~M. Mendoza, Y.~K. Kato, and D.~D. Awschalom.
\newblock Anisotropic interactions of a single spin and dark-spin spectroscopy
  in diamond.
\newblock {\em Nature Physics}, 1:94--98, 2005.

\bibitem{Hanzawa1993}
H.~Hanzawa, H.~Nishikori, Y.~Nisida, S.~Sato, T.~Nakashima, S.~Sasaki, , and
  N.~Miura.
\newblock Zeeman effect on the zero-phonon line of the nv center in synthetic
  diamond.
\newblock {\em Physica B: Condensed Matter}, 184:137 -- 140, 1993.

\bibitem{Manson2005}
N~B Manson and J~P Harrison.
\newblock Photo-ionization of the nitrogen-vacancy center in diamond.
\newblock {\em Diamond Relat. Mater.}, 14:1705--1710, October 2005.

\bibitem{Collins1983}
A~T Collins, M~F Thomaz, and M~I~B Jorge.
\newblock Luminescence decay time of the 1.945 ev centre in type ib diamond.
\newblock {\em J. Phys. C}, 16:2177 -- 2181, 1983.

\bibitem{Batalov2008}
A.~Batalov, C.~Zierl, T.~Gaebel, P.~Neumann, I.-Y. Chan, G.~Balasubramanian,
  P.~R. Hemmer, F.~Jelezko, and J.~Wrachtrup.
\newblock Temporal coherence of photons emitted by single nitrogen-vacancy
  defect centers in diamond using optical rabi-oscillations.
\newblock {\em Phys. Rev. Lett.}, 100(7):077401, 2008.

\bibitem{Zaitsev2001}
Alexander~M. Zaitsev.
\newblock {\em Optical Properties of Diamond: A Data Handbook}.
\newblock Springer, 2001.

\bibitem{Davies1976}
Gordon Davies and M~F Hamer.
\newblock Optical studies of the 1.945 ev vibronic band in diamond.
\newblock {\em Proc. R. Soc. Lond. A.}, 348:285--298, 1976.

\bibitem{Mohammed1982}
K~Mohammed, G~Davies, and A~T Collins.
\newblock Uniaxial stress splitting of photoluminescence transitions at optical
  centres in cubic crystals: theory and application to diamond.
\newblock {\em J. Phys. C}, 15:2779--2788, 1982.

\bibitem{Goss1996}
J.~P. Goss, R.~Jones, S.~J. Breuer, P.~R. Briddon, and S.~\"Oberg.
\newblock The twelve-line 1.682 ev luminescence center in diamond and the
  vacancy-silicon complex.
\newblock {\em Phys. Rev. Lett.}, 77(14):3041--3044, Sep 1996.

\bibitem{Gali2008}
Adam Gali, Maria Fyta, and Efthimios Kaxiras.
\newblock Ab initio supercell calculations on nitrogen-vacancy center in
  diamond: Electronic structure and hyperfine tensors.
\newblock {\em Phys. Rev. B}, 77:155206, 2008.

\bibitem{Davies1974}
Gordon Davies.
\newblock Vibronic spectra in diamond.
\newblock {\em J. Phys. C}, 7:3797--3809, 1974.

\end{thebibliography}

\end{document}